\documentclass[twoside,12pt]{article}
\usepackage{epsfig}

\newcommand{\be}{\begin{equation}}
\newcommand{\ee}{\end{equation}}
\newcommand{\bea}{\begin{eqnarray}}
\newcommand{\eea}{\end{eqnarray}}

\newcommand{\I}{\mathrm{i}}

\newcommand{\qbar}{\ensuremath{\overline{q}}}

\begin{document}

\title{ \vspace{1cm} Chiral symmetry breaking and the spin content of hadrons}
\author{L. Ya.\ Glozman, C. B.\ Lang, M.\ Limmer \\ 
\\
Institut f\"ur Physik, FB Theoretische Physik, Universit\"at Graz,
Universit\"atsplatz 5,\\
 A-8010 Graz, Austria \\
}
\maketitle
\begin{abstract}
From the parton distributions in the infinite momentum
frame one finds that only about 30\% of the nucleon spin
is carried by spins of the valence quarks, which gave rise
to the term "spin crisis". Similar results hold for the
lowest mesons, as it follows from the lattice simulations.
We define the spin content of a meson in the rest frame
and use a complete and orthogonal $\bar q q$ chiral basis
and a unitary transformation from the chiral basis to the
$^{2S+1}L_J$ basis. Then, given a mixture of different allowed
chiral representations in the meson wave function at a given 
resolution scale, one can obtain its spin content at this scale. 
To obtain the mixture of the chiral representations in the meson  
we measure in dynamical lattice simulations a ratio of couplings of
interpolarors with different chiral structure. For the $\rho$ meson
we obtain practically the $^3S_1$ state with no trace of the
spin crisis. Then a natural question arises:  which 
definition does reflect the spin content of a hadron?
\end{abstract}
\section{Introduction}

The nucleon spin structure function $g_1$, extracted from
the deep inelastic lepton-nucleon scattering with
polarization \cite{EMC1,EMC2}, if interpreted in the
infinite momentum frame, defines a fraction of
the nucleon spin carried by quark spins. Accordingly,
only about 30\% of the nucleon spin is carried by quark spins,
which is referred to usually as "spin crisis". Consequently,
a lot of experimental and theoretical efforts for the last 20 
years were devoted to search for  the rest of the
nucleon spin, without obvious success, however.  
Similar results are obtained on the lattice both
for nucleons and lowest mesons (for a review see ref. \cite{Hag}).

In  \cite{GLL1,GLL2,GLL3} we have suggested a way 
to define in a gauge invariant manner
and measure the spin content of mesons in the rest frame
at different resolution scales, including the infrared
ones. The method is based on the unitary transformation
from the complete and orthogonal chiral quark-antiquark
basis to the complete and orthogonal $^{2S+1}L_J$ basis
in the rest frame \cite{GN}. Using a set of interpolators
that transform according to different chiral representations
\cite{CJ,G1}, one can measure on the lattice a ratio of
couplings of different interpolators to a given state in
the rest frame. This ratio determines a mixture of different
chiral representations in the meson wave functions. Then,
given this ratio and using a unitary transformation from the
chiral basis to the angular momentum basis, one obtains the
angular momentum content of a meson in the rest frame.

The result was that the angular momentum content of the
$\rho$-meson in the rest frame is given approximately
by the $^3S_1$ wave, without obvious trace of the
spin crisis. Then a natural question is which of the
 definitions does reflect the spin content of a hadron?

Below we overview both principal aspects of the method
as well as the numerical results using as an example the
$\rho$-meson.

\section{The method}

 The $I=1,
J^{PC}=1^{--}$ $\bar q q$ states with unbroken chiral symmetry
transform according to two  chiral representations, $(0,1) +
(1,0)$ and $(1/2,1/2)_b$ \cite{CJ,G1}. These two representations 
form a complete
and orthogonal basis.
The  state that transforms as $(0,1) + (1,0)$ can be
created from the vacuum by the vector current, 
\begin{equation}
O_\rho^V(x)  = \qbar(x)\, \gamma^i \vec \tau\, q(x)\;,
\label{rV}
\end{equation}
and the state that belongs to the $(1/2,1/2)_b$ representation 
can be created by the pseudotensor operator,
\begin{equation}
O_\rho^T(x)  = \qbar(x)\, \sigma^{0i} \vec \tau\, q(x)\;.
\label{rT}
\end{equation}

In the continuum the physical $\rho$-meson with broken
chiral symmetry can be created from the vacuum by both
operators and the corresponding amplitudes are given as
\begin{eqnarray}
 \langle 0 | \qbar(0) \gamma^\mu q(0) | V(p; \lambda)\rangle &=& 
 m_\rho f_\rho^V e^\mu_\lambda\;,
\label{rhoV}\\
 \langle 0| \left(\qbar(0) \sigma^{\alpha \beta} q(0)\right)(\mu) | V(p; \lambda)\rangle &= &
 \I f_\rho^T(\mu) 
 (e^\alpha_\lambda p^\beta -  e^\beta_\lambda p^\alpha)\;,
\label{rhoT}
\end{eqnarray}
where   $V(p; \lambda)$ is the vector meson state with the mass $m_\rho$,
momentum $p$ and polarization $\lambda$. The vector current is conserved,
consequently the vector coupling constant $f_\rho^V$ is scale-independent. The
pseudotensor ``current'' is not conserved and is subject to a nonzero anomalous
dimension. Consequently the pseudotensor coupling $ f_\rho^T(\mu)$ manifestly
depends on the resolution scale $\mu$. 
The very fact that the $\rho$-meson can be created by the operators
that transform according two different chiral representations 
tells that chiral symmetry is broken  and the
meson wave function is a superposition of both chiral representations.

In the rest frame the ratio
\begin{equation} 
\frac{f_\rho^V}{f_\rho^T(\mu)} =  \frac
 {\langle 0 | \qbar(0) \gamma^i q(0) | V(\lambda)\rangle}
 {\langle 0 |\left(\qbar(0) \sigma^{0i} q(0)\right)(\mu) | V(\lambda)\rangle}
 \label{rhoV/rhoT}
\end{equation}
determines the ratio of the two allowed chiral
representations in the $\rho$-meson wave function
at a given resolution scale $\mu$.
Such  ratio can be measured on the  lattice.

Given a set of operators $O_i$ above we
use the variational method \cite{VAR} and 
calculate a cross-correlation matrix at zero spatial momentum  (i.e., in the
rest frame),
\begin{equation}\label{corr_inf}
C(t)_{ij}=\langle O_i(t)O_j^\dagger(0)\rangle=\sum_{n=1}^\infty a_i^{(n)} a_j^{(n)*} 
\mathrm{e}^{-E^{(n)} t}\;,
\end{equation}
with the coefficients giving the overlap of the  operators with the
physical state,
\begin{equation}\label{eq_w_f}
a_i^{(n)}=\langle 0| O_i|n\rangle\;.
\end{equation}

With a set of operators spanning a complete and orthogonal basis with  respect
to some symmetry group, these overlaps (coupling constants) give the complete
information about symmetry breaking. The interpolating composite operators $O_i$
are not normalized on the lattice
and consequently the absolute values of the coupling constants $a_i^{(n)}$
cannot be obtained. However, a ratio of the couplings is a well defined quantity
and can be computed as \cite{GLL1}
 \begin{equation}\label{ratio_op_comp}
\frac{a_i^{(n)}}{a_k^{(n)}}=
\frac{\widehat C(t)_{ij} u_j^{(n)}}{\widehat C(t)_{kj} u_j^{(n)}}\;.
\end{equation}
Here $\widehat C$ is the cross-correlation matrix from (\ref{corr_inf}), 
a  sum is implied for the index $j$ on the right-hand side and $u_j^{(n)}$ are the eigenvectors obtained from the generalized eigenvalue problem,
\begin{equation}\label{gev_1}
\widehat C(t)_{ij} u_j^{(n)} =\lambda^{(n)}(t,t_0)\widehat 
C(t_0)_{ij} u_j^{(n)}\;,
\end{equation}
with $t_0$ being some normalization point in Euclidean time. 
The ratio (\ref{ratio_op_comp})
coincides with the ratio of matrix elements
(\ref{rhoV/rhoT})  with $i \equiv
V; ~ k \equiv T$.

The chiral basis in the quark-antiquark system is a complete one and can be
connected to the complete angular momentum basis in the rest frame via the
unitary transformation \cite{GN}
\begin{equation}\label{unitary_1}
\left(
\begin{array}{l}
|(0,1)\oplus(1,0);1 ~ 1^{--}\rangle\cr
|(1/2,1/2)_b;1 ~ 1^{--}\rangle
\end{array}
\right) = U\cdot
\left(
\begin{array}{l}
|1;{}^3S_1\rangle\cr
|1;{}^3D_1\rangle
\end{array}
\right)
\end{equation}
with 
\begin{equation}\label{unitary_2}
U=
\left(
\begin{array}{cc}
\sqrt{\frac23} & \sqrt{\frac13} \cr
\sqrt{\frac13} & -\sqrt{\frac23} 
\end{array}
\right)\;.
\end{equation}
Consequently, if we know the mixture of the two allowed chiral representations in
a physical state, then applying the unitary transformation 
(\ref{unitary_1}) we are also able to obtain the angular momentum content of this
state in the rest frame. The mixture of the chiral representations
in the $\rho$-meson wave function at the resolution scale $\mu$ is given 
by (\ref{ratio_op_comp}) and (\ref{rhoV/rhoT}) and can be measured in
lattice simulations.

By this we define what is meant under the angular momentum content
of the $\rho$-meson in the rest frame  and, consequently, 
we can answer the question whether or not the spin of the $\rho$-meson
is carried by spins of its valence quarks in the rest frame. 

\section{Scale dependence of the chiral and angular momentum decompositions}

The ratio $a_V^{(n)}/a_T^{(n)}$ as well as a partial wave content of
a hadron are not the renormalization group invariant quantities. Hence
they manifestly depend on a resolution scale at which we probe the
hadron.
If we probe the hadron structure on the lattice with the local interpolators,
then we study the hadron decomposition at the scale fixed by
the lattice spacing $a$. For a reasonably small $a$ this scale
is close to the ultraviolet scale. However, we are interested 
in the hadron content at the infrared scales, where mass is
generated. For this purpose we cannot use a large $a$, because
matching with the continuum QCD will be lost. Given a fixed,
reasonably small lattice spacing $a$ a small resolution scale $\mu \sim 1/R$
can be achieved by the gauge-invariant smearing of the point-like
interpolators. We smear every quark field in spatial directions 
with the Gaussian profile over the size $R$ in
physical units such that $R/a\gg 1$, see Fig. 1. Then even in the continuum
limit $a \rightarrow 0$ we probe the hadron content at the resolution
scale fixed by $R$. Such a definition of the resolution is similar
to the experimental one, where an external probe is sensitive only to
quark fields (it is blind to gluonic fields) at a resolution 
that is determined by the momentum transfer in spatial directions.

\begin{figure}[tb]
 \begin{center}
 \includegraphics[width=.35\textwidth]{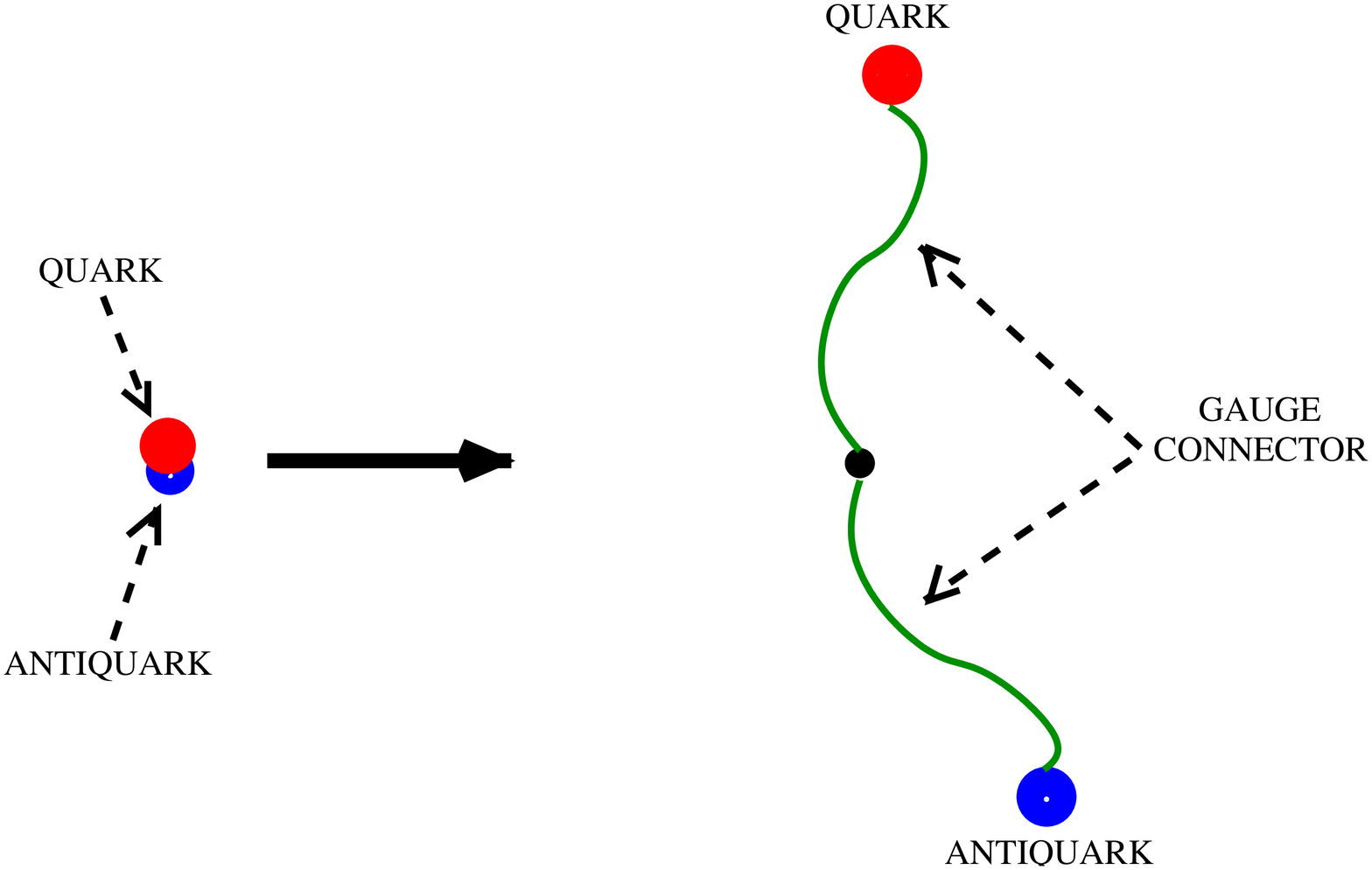}
 \includegraphics[width=.35\textwidth]{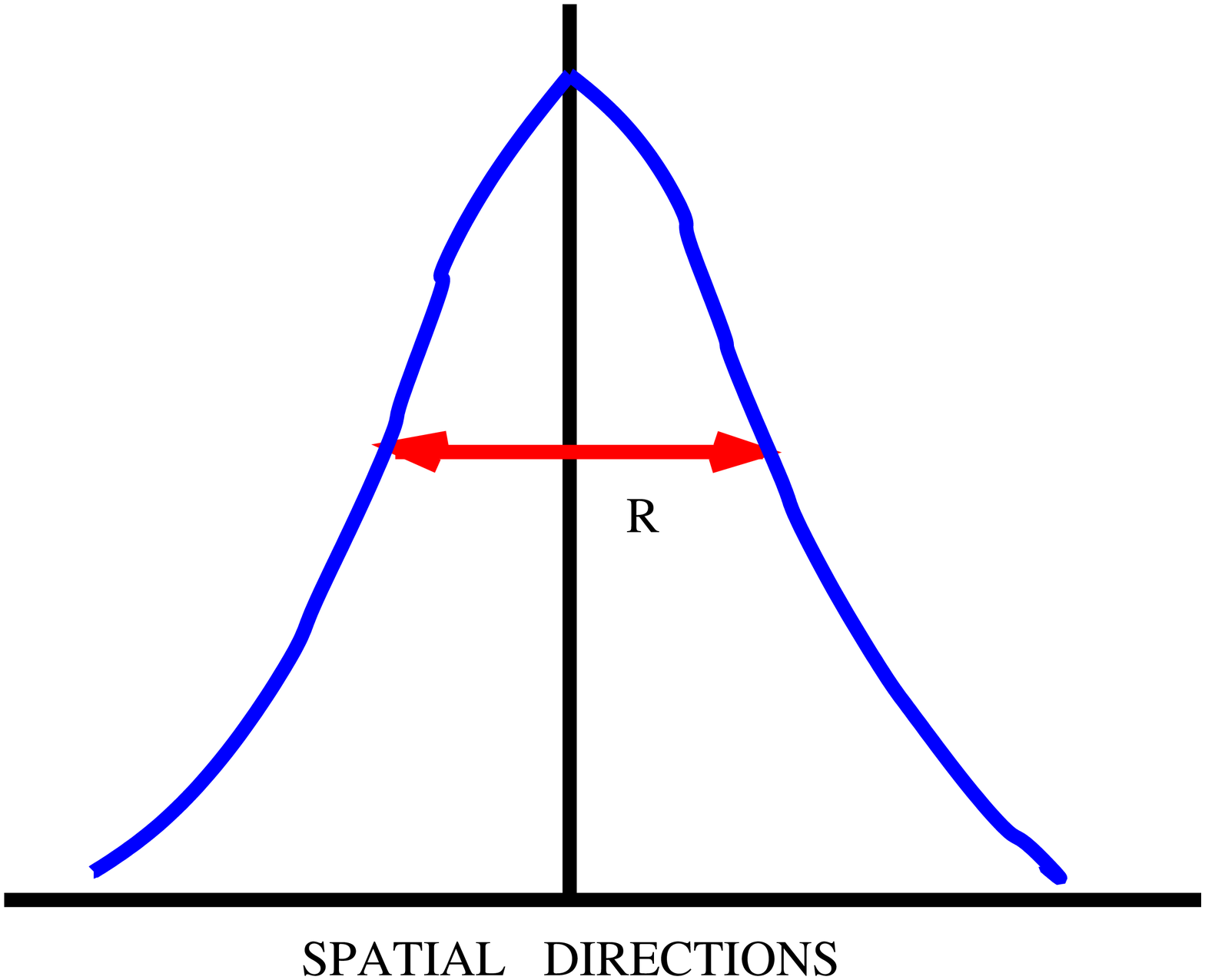}
 \end{center}
 \caption{Gauge-invariant smearing and the resolution scale $R$ definition}
\end{figure}

We use three different smearing radii R for the quark
fields both in the source and sink.
The ``narrow" smearing width (index $n$)  varies between
0.33 and 0.36 fm, depending on the set of configurations. The ``wide" smearing
radius (index $w$)  lies between 0.66 and 0.69 fm and the ``ultrawide" one 
is 0.81 - 0.85 fm (index $uw$).  Hence we can study the hadron structure at  resolutions 
0.33 fm  - 0.85 fm.

Consequently we have the following set of the interpolating operators 
\begin{eqnarray}
O^V_n=\overline u_n \gamma^i d_n\;,\;\;
&O^V_w=\overline u_w \gamma^i d_w\;,\;\;
&O^V_{uw}=\overline u_{uw} \gamma^i  d_{uw}\;,\;\;\nonumber\\
O^T_n=\overline u_n \gamma^t \gamma^i  d_n\;,\;\;
&O^T_w=\overline u_w \gamma^t \gamma^i  d_w\;,
&O^T_{uw}=\overline u_{uw} \gamma^t \gamma^i  d_{uw}\;,
\end{eqnarray}
where $\gamma^i$ is one of the spatial Dirac matrices and $\gamma_t$ is the
$\gamma$-matrix in (Euclidean) time direction.

\section{Lattice details and choices of correlation matrix}

\begin{table}
\begin{center}
\begin{tabular}{ccccccccc}
\hline
\hline
Set & $\beta_{LW}$ & $a\,m_0$ & \#{conf} & $a$ [fm] & $m_\pi$ [MeV] & $m_\rho$ [MeV] & $m_{\rho'}$ [MeV]\\
\hline
A\phantom{1} & 4.70 & -0.050 & 200 & 0.1507(17) & 526(7) & 911(11) & 1964(182)\\
B1           & 4.65 & -0.060 & 300 & 0.1500(12) & 469(5) & 870(10) & 1676(106)\\
B2           & 4.65 & -0.070 & 200 & 0.1406(11) & 296(6) & 819(18) & 1600(181)\\
C\phantom{1} & 4.58 & -0.077 & 300 & 0.1440(12) & 323(5) & 795(15) & 1580(159)\\
\hline
\hline
\end{tabular}
\caption{\label{tab:sim}
Specification of the data used here; for the gauge coupling only the
leading value $\beta_{LW}$ is given, $m_0$ denotes the bare mass parameter of
the CI action. Further details on the action, the simulation and the
determination of the lattice spacing and the $\pi$- and $\rho$-masses are found
in \cite{Gattringer:2008vj,Engel:2010my}.}
\end{center}
\end{table}

\begin{table}
\begin{center}
\begin{tabular}{cccc}
\hline
\hline
Set & $R_n$ [fm] & $R_w$ [fm] & $R_{uw}$ [fm]\\
\hline
A\phantom{1}  & 0.36 & 0.67 & --\\
B1            & 0.34 & 0.69 & 0.81\\
B2            & 0.34 & 0.66 & 0.85\\
C\phantom{1}  & 0.33 & 0.66 & --\\
\hline
\hline
\end{tabular}
\caption{\label{tab:radii}
Specification of the smearing radii $R$.}
\end{center}
\end{table}

\begin{figure}[t]
\begin{center}
\includegraphics[height=8cm,clip]{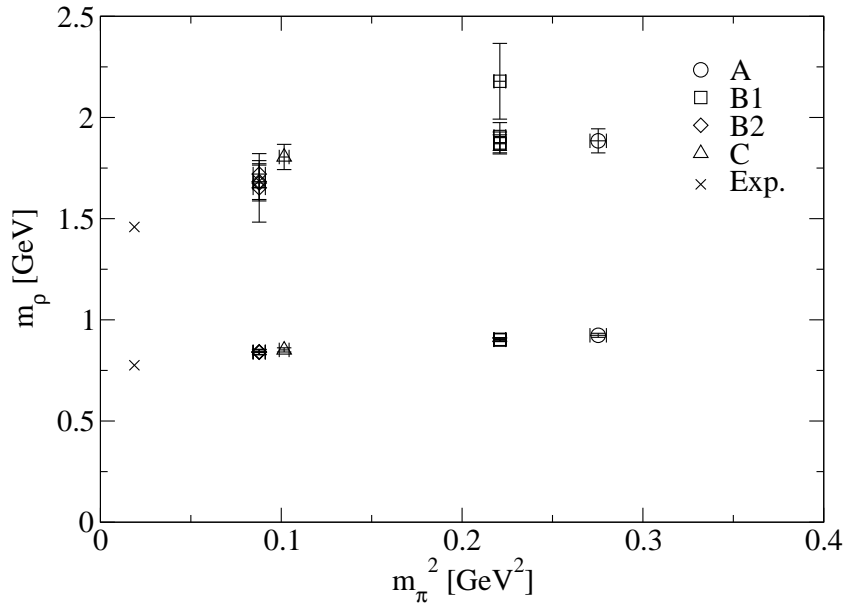}
\end{center}
\caption{\label{fig:masses}
The masses of both $\rho$ and $\rho'$ states extracted from
different $4 \times 4$ and $6 \times 6$ correlation matrices. The crosses indicate the mass
values from experiments.} 
\end{figure}

In our study we use Chirally Improved fermions \cite{CI} and the
L\"uscher-Weisz gauge action \cite{LW}. The lattice size is $16^3 \times 32$. We
use dynamical gauge configurations with two mass-degenerate light quarks.
With the lattice spacing $\approx 0.15$ fm the spatial volume of the  lattice is
$\approx 2.4^3$ fm$^3$.  In our present study we limit ourselves to the
$\rho$ and $\rho'$ states. For  details on the simulation we refer the reader to the
Table \ref{tab:sim} and to \cite{Gattringer:2008vj,Engel:2010my} .

For the sets A and C we
construct $4 \times 4$ correlation matrices (i.e., with both vector and
pseudotensor interpolators using narrow and wide smearing radii), while for
the sets B1 and B2 we study the $6 \times 6$ correlation matrix (with
narrow, wide and ultrawide smearings for both vector and pseudotensor
operators) as well as different  $4 \times 4$ sub-matrices.

In Fig. \ref{fig:masses} we show masses of both the ground state
$\rho$-meson and its first excitation $\rho'$ extracted from different
correlation matrices.

\section{Chiral symmetry breaking and the angular momentum content
of $\rho$ and $\rho'$ mesons}

\begin{figure}[t]
\begin{center}
\includegraphics*[height=8cm,clip]{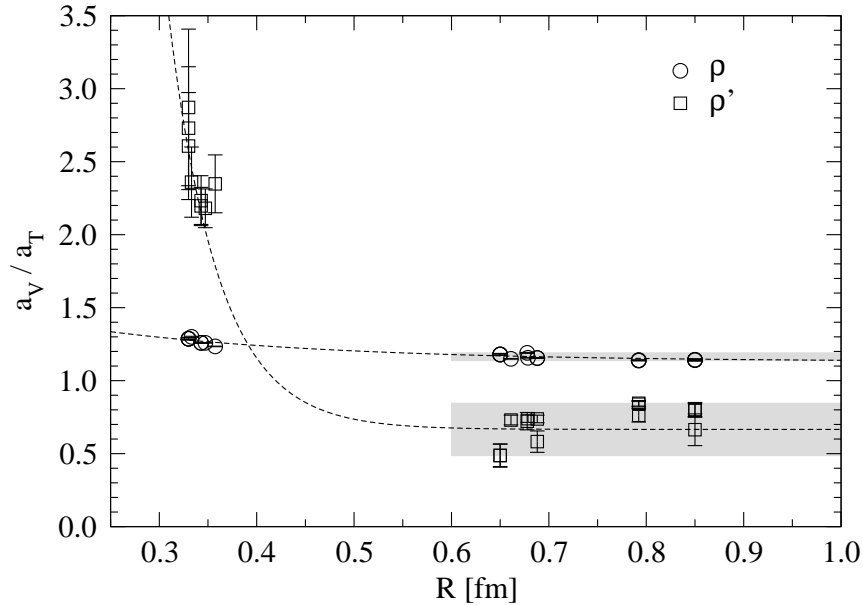}
\end{center}
\caption{\label{fig:all}
A ratio of the vector to the pseudotensor couplings versus a resolution scale $R$, as
extracted from all  $4 \times 4$ and $6 \times 6$ correlation matrices.
Broken lines are drawn only to guide the eye.}
\end{figure}

The ratio $a_V/a_T$
of the two allowed chiral representations in the $\rho$- and
$\rho'$-mesons versus the resolution $R$ is shown on Fig. \ref{fig:all}.
The results obtained from different $4 \times 4$ and $6 \times 6$
correlation matrices are consistent with each other. For the
ground state $\rho$ we observe strong chiral symmetry breaking
practically at all resolutions; only in the deep ultraviolet $R,a \ll 0.3$,
where chiral symmetry is unbroken, the tensor "current" decouples
from the $\rho$-meson. For the $\rho'$ the tensor operator decouples
much faster towards the ultraviolet than for the $\rho$-meson.
This shows that the wave functions of these two states are
significantly different and the chiral symmetry breaking is
different in both states. At the scale of $\sim 1$ fm
of the hadron size, i.e., at which the hadron mass is generated,
chiral symmetry breaking in the $\rho'$ state is less essential
than in the ground state.

Given this ratio and the unitary transformation from the chiral basis
to the angular momentum basis (\ref{unitary_1}) one can obtain
the partial wave content of both states.
 For
the $\rho$ meson it is approximately  $0.99\,|^3S_1\rangle -
0.1\,|^3D_1\rangle$. Hence the ground state in the infrared is practically a
pure $^3S_1$ state with a tiny admixture of the $^3D_1$ wave.
Consequently in the rest frame the spin of the $\rho$-meson 
is almost completely carried by spins of its valence quarks. No
trace of the "spin crisis" is observed. In this our definition
we do not separate contributions of quarks and gluons into
hadron spin. In the confining regime, where the gauge invariance 
and the quark-gluon interaction are
of crucial importance, it is not possible to separate both contributions
into total spin in a sensible gauge-invariant manner; after all the
 quark-gluon interaction is a consequence of the gauge invariance. 
The spin of the $\rho$-meson in the rest frame is carried by spins of its 
valence quarks dressed by gluons.
The gluonic field is important for the angular momentum generation, 
because it is this field that
provides chiral symmetry breaking and that is responsible for most of the 
hadron mass. 

In the excited $\rho$-meson there
is a significant contribution of the $^3D_1$ wave. In the
latter case the angular momentum content is between the following 
two lower and
upper bound values. For the lower bound it is $0.88\,|^3S_1\rangle -
0.48\,|^3D_1\rangle$ and for the  upper bound it is $0.97\,|^3S_1\rangle -
0.25\,|^3D_1\rangle$. This once again demonstrates that the first 
excitation of
the $\rho$-meson cannot be considered  a pure radial excitation 
of the ground state $\rho$. Obviously,
both radial and orbital degrees of freedom are excited which reflects yet
unknown dynamics of confinement and chiral symmetry breaking.

\paragraph{Acknowledgments.}

We gratefully acknowledge support of the grants P21970-N16 and DK W1203-N08 
of the Austrian
Science Fund FWF and the DFG project SFB/TR-55. The calculations have been performed
on the SGI Altix 4700 of the Leibniz-Rechenzentrum Munich and on the local
clusters at the University of Graz.

\end{document}